\documentstyle[12pt]{article}
\if@twoside\oddsidemargin25mm
\else\oddsidemargin25mm\evensidemargin25mm\marginparwidth25mm\fi%
\def\thefootnote{\dag}
\def\bl{\Big\{}
\def\br{\Big\}}
\def\bpl{\Big(}
\def\bpr{\Big)}
\def\ve{\varepsilon}

\def\t{\theta}
\def\vphi{\varphi}

\def\O{\Omega}
\def\F{{\cal F}}
\def\hf{\frac{1}{2}}
\def\der{\partial}
\def\bq{\begin{equation}}
\def\eq{\end{equation}}
\def\brr{\begin{eqnarray}}
\def\err{\end{eqnarray}}
\def\ba{\left(\begin{array}}
\def\ea{\end{array}\right)}

\def\Dslash{\hbox{\ooalign{$\displaystyle D$\cr$\hspace{.03in}/$}}}

\def\Vslash{\hbox{\ooalign{$\displaystyle V$\cr$\hspace{.02in}/$}}}
\begin{document}
% new commands
\newcommand{\dr}{\raise.3ex\hbox{$\stackrel{\leftarrow}{\partial }$}{}}
\newcommand{\dl}{\raise.3ex\hbox{$\stackrel{\rightarrow}{\partial}$}{}}
\newcommand{\eqn}[1]{(\ref{#1})}
\newcommand{\ft}[2]{{\textstyle\frac{#1}{#2}}}
\newcommand{\dkt}{\delta _{KT}}
\newcommand{\QED}{{\hspace*{\fill}\rule{2mm}{2mm}\linebreak}}
\renewcommand{\theequation}{\arabic{equation}}
\csname @addtoreset\endcsname{equation}{section}
\newsavebox{\uuunit}
\sbox{\uuunit}
    {\setlength{\unitlength}{0.825em}
     \begin{picture}(0.6,0.7)
        \thinlines
        \put(0,0){\line(1,0){0.5}}
        \put(0.15,0){\line(0,1){0.7}}
        \put(0.35,0){\line(0,1){0.8}}
       \multiput(0.3,0.8)(-0.04,-0.02){12}{\rule{0.5pt}{0.5pt}}
     \end {picture}}
\newcommand {\unity}{\mathord{\!\usebox{\uuunit}}}
\newcommand  {\Rbar} {{\mbox{\rm$\mbox{I}\!\mbox{R}$}}}
\newcommand  {\Hbar} {{\mbox{\rm$\mbox{I}\!\mbox{H}$}}}
\newcommand {\Cbar}
    {\mathord{\setlength{\unitlength}{1em}
     \begin{picture}(0.6,0.7)(-0.1,0)
        \put(-0.1,0){\rm C}
        \thicklines
        \put(0.2,0.05){\line(0,1){0.55}}
     \end {picture}}}
\newsavebox{\zzzbar}
\sbox{\zzzbar}
  {\setlength{\unitlength}{0.9em}
  \begin{picture}(0.6,0.7)
  \thinlines
  \put(0,0){\line(1,0){0.6}}
  \put(0,0.75){\line(1,0){0.575}}
  \multiput(0,0)(0.0125,0.025){30}{\rule{0.3pt}{0.3pt}}
  \multiput(0.2,0)(0.0125,0.025){30}{\rule{0.3pt}{0.3pt}}
  \put(0,0.75){\line(0,-1){0.15}}
  \put(0.015,0.75){\line(0,-1){0.1}}
  \put(0.03,0.75){\line(0,-1){0.075}}
  \put(0.045,0.75){\line(0,-1){0.05}}
  \put(0.05,0.75){\line(0,-1){0.025}}
  \put(0.6,0){\line(0,1){0.15}}
  \put(0.585,0){\line(0,1){0.1}}
  \put(0.57,0){\line(0,1){0.075}}
  \put(0.555,0){\line(0,1){0.05}}
  \put(0.55,0){\line(0,1){0.025}}
  \end{picture}}
\newcommand{\Zbar}{\mathord{\!{\usebox{\zzzbar}}}}
\newcommand{\Ka}{K\"ahler}
\newcommand{\qu}{quaternionic}
\def\ib{{\bar \imath}}
\def\jb{{\bar \jmath}}
\renewcommand{\sp}{{Sp\left( 2n+2,\Rbar\right)}}
%%%%%%
\renewcommand{\a}{\alpha}
\renewcommand{\b}{\beta}
\renewcommand{\c}{\gamma}
\renewcommand{\d}{\delta}
\newcommand{\pa}{\partial}
\newcommand{\g}{\gamma}
\newcommand{\G}{\Gamma}
\newcommand{\A}{\Alpha}
\newcommand{\B}{\Beta}
\newcommand{\D}{\Delta}
\newcommand{\e}{\epsilon}
\newcommand{\E}{\Epsilon}
\newcommand{\z}{\zeta}
\newcommand{\Z}{\Zeta}
\newcommand{\K}{\Kappa}
\renewcommand{\l}{\lambda}
\renewcommand{\L}{\Lambda}
\newcommand{\m}{\mu}
\newcommand{\M}{\Mu}
\newcommand{\N}{\Nu}
\newcommand{\x}{\chi}
\newcommand{\X}{\Chi}
\newcommand{\p}{\pi}
\newcommand{\R}{\Rho}
\newcommand{\s}{\sigma}
\renewcommand{\S}{\Sigma}
\newcommand{\T}{\Tau}
\newcommand{\y}{\upsilon}
\newcommand{\Y}{\upsilon}
\renewcommand{\o}{\omega}
\newcommand{\q}{\theta}
\newcommand{\h}{\eta}
\begin{titlepage}
\begin{flushright} KUL-TF-95/38\\ THU-95/35\\ ULB-TH-95/16 \\
hep-th/9512143
\end{flushright}
\vfill
\begin{center}{\LARGE\bf The Vector-Tensor Supermultiplet \\[1.5mm]
with Gauged Central Charge}\\
\vspace{2cm}
{\large P. Claus$^1$\footnote{Wetenschappelijk Medewerker, NFWO, Belgium},
        B. de Wit$^2$, M. Faux$^2$\\[1.5mm]
        B. Kleijn$^2$, R. Siebelink$^3$ and P. Termonia$^1$}\\
\vspace{7mm}
{\small
$^1$ Instituut voor Theoretische Fysica - Katholieke Universiteit Leuven\\
     Celestijnenlaan 200D B--3001 Leuven, Belgium\\[6pt]
$^2$ Institute for Theoretical Physics - Utrecht University\\
     Princetonplein 5, 3508 TA Utrecht, The Netherlands\\[6pt]
$^3$ Service de Physique Th\'{e}orique, Universit\'{e} Libre de
     Bruxelles,\\
     Campus Plaine, CP 225, Bd du Triomphe,\\B--1050 Bruxelles, Belgium}
\end{center}
\vfill
\begin{center} {\bf Abstract}\end{center}
{\small
The vector-tensor multiplet is coupled off-shell to an $N=2$ vector
multiplet such that its central charge transformations are
realized locally. A gauged central charge is a necessary
prerequisite for a coupling to supergravity and the strategy
underlying our construction uses the potential for such a
coupling as a guiding principle. The results for
the action and transformation rules take a nonlinear form
and necessarily include a Chern-Simons term. After a duality
transformation the action is encoded in a homogeneous holomorphic function
consistent with special geometry.}
\vspace{7mm}
\flushleft{December 1995}
\end{titlepage}

\renewcommand\thefootnote{\arabic{footnote}}
\setcounter{footnote}{0}
\noindent
Off-shell $N=2$ supermultiplets have at least eight bosonic and
eight fermionic components. The smallest of these constitute a
variety with two distinct subsets. In the first subset are two
multiplets which on shell describe one spin-1, two spin-0, and a doublet of
spin-1/2 massless states. These are the {\it vector multiplet}
\cite{grimm}, which includes
a complex scalar, a vector gauge field and a triplet of auxiliary
scalars, and the {\it vector-tensor multiplet}
\cite{sohnius,DWKLL},  with
a real scalar, a vector gauge field, a tensor gauge field, and a
real auxiliary scalar.  In the second subset are three multiplets
which on shell describe four spin-0 and a doublet of spin-1/2 states.
These are the {\it hypermultiplet} \cite{fayet}, with four real
scalars and four real auxiliary scalars,
the {\it tensor multiplet} \cite{DWVH}, with a triplet of scalars,
a tensor gauge field and a complex auxiliary scalar, and the {\it
double-tensor multiplet}, with two real scalars and two
tensor gauge fields.  Within a given subset, the alternative
field-theoretic formulations are equivalent
on-shell in the sense that their linearized field equations lead to the
same states. They are, however, inequivalent off-shell. Unlike in
$N=1$ supersymmetry, it is not possible to convert one field
representation into another in a way that leaves the full $N=2$
supersymmetry manifest. This aspect is presumably tied to the
presence of an off-shell central charge which is required for all
of these multiplets other than the vector and the tensor multiplet.
In terms of $N=1$ supersymmetry, the conversion between
different multiplets involves the replacement of a chiral by a tensor
(linear) supermultiplet or vice versa.

The off-shell features of these multiplets are crucial
for understanding their general couplings. This is directly
related to vector and tensor gauge invariances that must be
preserved. In the context of local supersymmetry, there is
a further restriction since the central charge must also be
associated with a local symmetry. These aspects are
important when considering string compactifications, where the
axion field emerges {}from a tensor gauge field. In the $N=2$
effective action this tensor field must be part of an $N=2$
supermultiplet\footnote{For a recent discussion
of $N=2$ dilaton assignments, see\cite{siegel}}.
For the heterotic string the
dilaton-axion complex is contained in a vector-tensor multiplet,
which for practical reasons is often converted into a vector
multiplet. For type-IIA strings it is contained in a tensor multiplet and
for type-IIB in a double-tensor multiplet, either of which can
be converted to a hypermultiplet. To understand the systematics of
the various couplings of the dilaton-axion complex, it seems
advantageous to consider that multiplet which is most closely
related to the vertex operators in the underlying string theory
in order to fully exploit the restrictions at the level of the
effective action, especially when one considers data beyond the
spectrum and the Yukawa couplings.

The relevance of these issues is two-fold. Because the dilaton acts as the
loop-counting parameter in string perturbation theory, its
generic couplings have a direct bearing on the perturbative
features of string theory. In this context the $N=2$ nonrenormalization
theorems play an important role. Then there are subtle relations
amongst string groundstates, through mirror symmetry and through
string-string duality, which are of truly
nonperturbative nature. These
relations are also described at the level of
four-dimensional $N=2$ supersymmetric effective actions, where
the assignment of the dilaton-axion complex to an appropriate
supermultiplet forms a crucial ingredient. This is the
motivation for the work described in this letter, where we
consider the coupling of the vector-tensor multiplet to an $N=2$ vector
multiplet background that is associated with {\it local} central
charge transformations. Local central charge transformations are
necessary in supergravity, as was first exhibited in
\cite{zachos} for massive hypermultiplets and in \cite {DWVHVP}
for off-shell hypermultiplets in the context of conformal
supergravity. We base ourselves on
the linearized transformation rules given in \cite{DWKLL},
{}from which it is already clear that
the gauge field of the central charge transformations cannot
be the vector field of the vector-tensor multiplet
itself. As a guiding principle for deriving the coupling we
require the background fields to couple such that we can
consistently assign the multiplet components to a representation of
conformal supersymmetry. In this way we hope to incorporate all
the crucial features necessary for a coupling to supergravity by
means of the superconformal multiplet calculus used in the past.

The vector-tensor multiplet contains
a scalar field $\phi$, a vector gauge field $V_\mu$, a tensor gauge field
$B_{\mu\nu}$ and a doublet of Majorana spinors $\lambda_i$.
As mentioned above, the multiplet includes a central charge.
The vector multiplet background contains a complex scalar $X$,
a spinor doublet $\O_i$, the gauge field $W_\mu$, now associated
with the central charge, and the auxiliary fields $Y_{ij}$
\footnote{We use the chiral notation employed in
  \cite{DWVHVP,DWLVP}, where, for spinor quantities, upper and
  lower $SU(2)$ indices $i,j,\ldots$ denote chiral components.
  For the spinors used in this letter, the positive chirality
  spinors are $\O_i$, $\lambda_i$ and $\e^i$ and thus satisfy
  $\g_5\O_i=\O_i$, etc.
  The $SU(2)$ indices are raised and lowered by complex
  conjugation. Antisymmetrization is defined with ``weight one",
  so that, for example,
  $\der_{[\mu}V_{\nu]}=\hf(\der_\mu V_\nu-\der_\nu V_\mu)$,
  and similarly for symmetrization.
  A dual tensor is defined by
  $\tilde{F}^{\mu\nu}=\hf\ve^{\mu\nu\rho\s}F_{\rho\s}$ with
  $\ve^{1234}=1$, and chiral components are defined by
  $F_{\mu\nu}^\pm=\hf(F_{\mu\nu}\pm \tilde{F}_{\mu\nu})$.}.
The supersymmetry transformations, the central
charge, and two additional gauge invariances describe an unusual
geometry which requires explanation.
We begin by discussing the central charge.  Infinitesimally,
this acts as $\d_z\phi=z\phi^{(z)}$.
Successive applications generate a sequence of translations,
\bq \phi\longrightarrow\phi^{(z)}\longrightarrow\phi^{(zz)}
    \longrightarrow {\rm etc}, \label{hierarchy}
 \eq
and similarly on the remaining fields.
As is well known, such a hierarchy arises naturally when starting
{}from a five-dimensional supersymmetric theory with one
compactified coordinate,
but this interpretation is not essential here.
The field $\phi^{(z)}$ is an independent scalar. In contrast, all
other objects in the hierarchy, $\phi^{(zz)}, V_{\mu}^{(z)},
V_\mu^{(zz)}$, etcetera,
are dependent, and are given by particular combinations of the
independent fields.  This is enforced by a set of constraints, which
we exhibit below. The two gauge transformations include
a tensor transformation with parameter $\Lambda_\mu$,
under which $B_{\mu\nu}\rightarrow B_{\mu\nu}+\der_{[\mu}\Lambda_{\nu]}$,
and a vector transformation, with parameter $\t$,
under which $V_\mu\rightarrow V_\mu+\der_\mu\t$.
Closure of the algebra requires that $B_{\mu\nu}$ transform as well under
the vector gauge transformation and couple
to a Chern-Simons form.  The transformation rules are
determined by imposing closure of the algebra.
Modulo field redefinitions these rules are given by
\brr \d W_\mu &=& \der_\mu z \,,\nonumber\\
     \d V_\mu &=& \der_\mu\t+z\, V^{(z)}_\mu \,,\nonumber\\
     \d B_{\mu\nu} &=& \der_{[\mu}\L_{\nu]}+\t\,\der_{[\mu}V_{\nu]}
     +z \,B_{\mu\nu}^{(z)}\,,
 \label{gauge}\err
where $\theta$, $\L_\m$ and $z$ are spacetime-dependent parameters.
Note that at the linearized level, the need for the Chern-Simons
modification is not apparent since a transformation
$\d B_{\mu\nu}\propto\der_{[\mu}V_{\nu]}$ can be regarded as a
field-dependent tensor gauge transformation.
Imposing closure of the algebra on $V_\mu$, one readily concludes that
$V_\mu^{(z)}$ is invariant under the $\theta$ transformation
and also that $B_{\mu\nu}^{(z)}$ takes the following form,
\bq B_{\mu\nu}^{(z)}=\hat B_{\mu\nu}^{(z)}
    -V_{[\mu}V_{\nu]}^{(z)}\,,
 \label{bzdef}\eq
where $\hat B_{\mu\nu}^{(z)}$ is invariant under the
$\theta$ transformation. The $z$ and $\theta$ gauge transformations do not
commute, but close into a tensor gauge transformation with parameter
$\Lambda_\m\propto z\,\theta\,V_\m^{(z)}$.
Derivatives covariant with respect to
central charge transformations are given by
$D_\mu\phi=\der_\mu\phi-W_\mu\phi^{(z)}$, for instance.  The field
strengths are
\brr \F_{\mu\nu} &=& 2 \der_{[\mu}W_{\nu]}\,, \nonumber\\
     F_{\mu\nu} &=& 2 \der_{[\mu}V_{\nu]}-2W_{[\mu}V_{\nu]}^{(z)}\,,
     \nonumber\\
     H^\mu &=& \ft{i}{2}\ve^{\mu\nu\l\s}
     \bpl \der_\nu B_{\l\s}-V_\nu\der_\l V_\s
     -W_\nu\hat{B}_{\l\s}^{(z)}\bpr\,,
 \label{fieldstr}\err
where $\hat{B}_{\mu\nu}^{(z)}$ is defined in (\ref{bzdef}).
These field strengths are invariant
under vector and tensor gauge transformations.  Under a
central charge transformation, $\F_{\mu\nu}$ is invariant
and both $F_{\mu\nu}$ and $H^\mu$ are covariant.
The field strengths satisfy the following Bianchi identities
\brr D_\mu\tilde{F}^{\mu\nu} &=& -V_\mu^{(z)}\tilde\F^{\mu\nu}\,,\nonumber\\
     D_\mu H^\mu &=& -\ft{i}{4}F_{\mu\nu}\tilde F^{\mu\nu}
     -\ft{i}{2} \hat B_{\mu\nu}^{(z)}\tilde\F^{\mu\nu}\,.
 \label{Bianchi}\err
Notice the appearance of $F_{\mu\nu}\tilde{F}^{\mu\nu}$ in the
Bianchi identity for $H^\mu$.  This is related to the
the Chern-Simons form $V_{[\mu}\der_\nu V_{\l]}$ appearing in the
definition of $H^\mu$ in equation (\ref{fieldstr}), which was also
mentioned previously.
This coupling is unavoidable if the algebra is to close
in the presence of the vector multiplet background,
and is responsible for significant nonlinearities in the supersymmetry
transformation rules as we will discuss.

The vector-tensor multiplet is an assembly of an $N=1$
vector multiplet and an $N=1$ tensor multiplet. Since these
multiplets carry different conformal weights,
it is impossible to assemble them
directly into an $N=2$ representation of the superconformal
algebra. This incompatiblity could have
been anticipated by observing
that kinetic terms for vector and tensor gauge fields are
not conformally invariant in the same spacetime dimension.
However, the vector multiplet, which provides the gauge field
required to gauge the central charge, can simultaneously provide
fields to compensate for the difference in conformal weights
\cite{DWVHVP}. Hence this vector
multiplet plays a dual role; it provides the gauge field
for the central charge and it also enables us to construct
transformation rules which are covariant with respect to
the bosonic part of the superconformal algebra.  These
modifications compensate for lack of covariance with respect to
both scale and chiral transformations. The structure is therefore
constrained. In this way we find an extra bonus because the
central charge transformations
in the algebra become field-dependent so as to become central
with respect to the full $N=2$ superconformal algebra.

If the scalar field of the background vector multiplet is set to
a constant and the other components are set to zero,
the full supersymmetry is retained,
but scale and chiral transformations are broken.
In this limit, one obtains a vector-tensor multiplet with
modifications which are nonlinear in the vector-tensor fields. One
of these modifications is the coupling of the tensor field
to the Chern-Simons form. There is a singular limit in which these
nonlinear modifications disappear.

The vector multiplet is completely fixed as a superconformal
multiplet\footnote{The Weyl and chiral weights for most
multiplets are summarized in the first paper in \cite{DWLVP}}.
For instance, $X$ transforms under dilatations with
weight $w=1$ and under chiral $U(1)$ transformations with weight
$c=-1$. The vector-tensor multiplet is not so restricted.
For instance, by multiplying by powers of $|X|$ we can
adjust the conformal weight of $\phi$ arbitrarily.
We use this freedom to choose that $\phi$ transforms
under dilatations with weight $w=0$
(the chiral weight of $\phi$ must be zero since it is a real field).
In a similar manner we adjust the weights of $\l_i$ to
$w=c=1/2$. The gauge fields $V_\m$ and $B_{\m\nu}$ must have
$w=c=0$; any other assignment would create an obstruction
between their corresponding gauge transformations and local scale
and chiral transformations.  In
this letter we omit further details of how we obtained our results.
Instead, we simply present the various nonlinear constraints,
the supersymmetry transformation rules,
and briefly discuss the supersymmetric action and its symmetries.
We defer a deeper discussion to a
more comprehensive presentation which is forthcoming.

The central charge commutes with all other
transformations up to gauge transformations.
Therefore by successive application of the central charge one generates an
infinite hierarchy of vector-tensor multiplets, as already
indicated in (\ref{hierarchy}), whose components have the same weights and
have the same transformation rules, modulo gauge
transformations.  However, the components of these new multiplets
are not independent. At the same time one has an infinite set
of constraints, required by the closure of the supersymmetry algebra,
so that the vector-tensor multiplet has precisely eight bosonic
and eight fermionic components.  Therefore, with the exception of
$\phi^{(z)}$, there are no additional degrees of freedom.
The constraints can be concisely summarized by giving the expressions for
$V_\mu^{(z)}, B_{\mu\nu}^{(z)}, \l_i^{(z)}$ and $\phi^{(zz)}$,
{}from which all other constraints follow by successively applying
central charge transformations,
\brr V_\mu^{(z)} &=& \frac{-1}{4\phi |X|^2} H_\mu+\frac{i\phi}{2}
     \der_\mu\ln\frac{\bar X}{X} \nonumber\\
     & &+ \frac{i}{2\phi|X|^2} \bl \bpl \bar X\bar \l^i + {\phi}
     \bar \O^i \bpr \g_\mu \bpl X\l_i + {\phi} \O_i\bpr
     -\ft12\phi^2 \bar \O^i \g_\mu \O_i\br\,,\nonumber\\
     %%%
     B_{\mu\nu}^{(z)} &=& -V_{[\mu}V_{\nu]}^{(z)}
     +i\phi\tilde F_{\mu\nu}
     -\ft12 {\phi^2}{\cal F}_{\mu\nu}  -\ve^{ij} X \bar \l_i
     \s_{\mu\nu} \l_j
     -\ve_{ij} \bar X \bar \l^i \s_{\mu\nu} \l^j\,,\nonumber\\
     %%%
     \l_i^{(z)} &=& \frac{-1}{4|X|^2}\ve_{ij}\Dslash\bpl
     2\bar{X}\l^j+\phi\O^j\bpr
     -\frac{1}{4\phi X}\ve_{ij}\bpl\Dslash\phi\bpr\l^j \nonumber\\
     & & -\frac{i}{8\phi|X|^2}\bl\s\cdot\bpl F-i\phi\F\bpr\l_i
     -i\phi\ve_{ik}Y^{kj}\l_j\br\nonumber\\
     & & -\frac{\phi^{(z)}}{2\phi X}\bpl X\l_i+2\phi\O_i\bpr
     -\frac{i}{4|X|^2\phi}\Vslash^{(z)}
     \ve_{ij}\bpl\bar{X}\l^j+\phi\O^j\bpr
     +\mbox{ fermion trilinears}\,,\nonumber\\
     %%%
     \phi^{(zz)}&=&- \frac{1}{8\phi|X|^2}(D_\mu\phi)^2 -
     \frac{1}{4|X|^2}D^2\phi-\frac{1}{2|X|^2}D_\mu \phi\; \partial^\mu
     \ln|X| \nonumber\\
     & & +\frac{\phi}{8|X|^6}\bl\big(X\der_\mu
     \bar{X} -\bar X \der_\m X\big)^2
     -|X|^2\big(X\Box \bar{X}-\bar X\Box{X}\big)\br  \nonumber\\
     & & +\frac{1}{128 \phi^3 |X|^6} \bpl H_\mu + 2i\phi^2|X|^2
     \partial_\mu \ln\frac{\bar X}{X}\bpr^2
     +\frac{1}{64|X|^4\phi}\big(F_{\mu\nu}+i\phi\tilde{\F}_{\mu\nu}\big)^2
     \nonumber\\
     & &-\frac{\big(\phi^{(z)}\big)^2}{2\phi}
     +\frac{\phi}{64|X|^4}Y_{ij}Y^{ij}  +\mbox{ fermion terms}.
 \label{central}\err
With these constraints the supersymmetry algebra closes
upon anticommutation into
a spacetime translation, a vector and a tensor gauge
transformation, and a central charge transformation.
One may notice that the $B_{\mu\nu}^{(z)}$ equation is not invariant
under a vector gauge transformation. This
equation can, however, be cast in a $\t$-invariant form by expressing
it in terms of $\hat{B}_{\mu\nu}^{(z)}$, defined in (\ref{bzdef}).

The supersymmetry transformation rules for the independent
fields, with all of the nonlinear modifications discussed above, are
\brr \d\phi &=& \bar{\e}^i\l_i
     +\bar{\e}_i\l^i \,,\nonumber\\
     %%%
     \d V_\mu &=& i\ve^{ij}\bar{\e}_i\g_\mu
     \bpl 2X\l_j+\phi\O_j\bpr
     -iW_\mu\bar{\e}^i\l_i + h.c. \,, \nonumber\\
     %%%
     \d B_{\mu\nu} &=& -\bar{\e}^i\s_{\mu\nu}
     \bpl 8\phi |X|^2 \l_i+4 \phi^2 \bar X \O_i\bpr
     -i\ve^{ij}\bar{\e}_i\g_{[\mu}V_{\nu]}
     \bpl 2X\l_j+\phi\O_j\bpr  \nonumber\\
     & & +\ve^{ij}\bar{\e}_i\g_{[\mu}W_{\nu ]}
     \bpl 4\phi X\l_j+\phi^2 \O_j\bpr
     -i \bar{\e}^i \l_i\, W_{[\mu}V_{\nu]}  +h.c.\,, \nonumber\\
     %%%
     \d\l_i &=& \bpl\Dslash\phi-i\Vslash^{(z)}\bpr\e_i
     -\frac{i}{2X}\ve_{ij}\s\cdot\big(F-i\phi\F\big)\e^j
     +\big(2\ve_{ij}\bar{X}\phi^{(z)}
     -\frac{\phi}{2X}Y_{ij}\big)\e^j\nonumber\\
     & & +\frac{1}{X}\bl\frac{1}{2\phi}\e^j\big(X\,\bar\l_i\l_j
     -\bar{X}\,\ve_{ik}\ve_{jl}\bar\l^k\l^l\big)
     -\big(\bar\e^j\O_j\big)\l_i-\big(\bar\e^j\l_j\big)\O_i\br\,.
 \err
Observe that these transformation rules are not linear in the
vector-tensor fields; $\d B_{\m\nu}$ is quadratic in these fields
and also $\d \l_i$ contains quadratic
terms. This nonlinearity is linked to the Chern-Simons modification
discussed above.  We stress that these nonlinearities
are unavoidable if the vector-tensor multiplet is to exist
in a superconformal background.

The results presented so far have
analogs at all higher levels in the central charge.
To elucidate the structure more completely we introduce a notation
where, for positive integers ${\rm Z}$, we have
${\cal O}^{({\rm Z})}=({\cal O}^{(z)},
{\cal O}^{(zz)},...)$, so that
\bq\d_z{\cal O}^{({\rm Z})}=z\,{\cal O}^{({\rm Z}+1)}.
 \eq
Covariant derivatives are given by $D_\mu{\cal O}^{({\rm Z})}=
\der_\mu{\cal O}^{({\rm Z})}-W_\mu{\cal O}^{({\rm Z}+1)}$.
The objects $\phi^{({\rm Z})}, V_\mu^{({\rm Z})}$, and $\l_i^{({\rm Z})}$ are
$\t$-invariant, but $B_{\mu\nu}^{({\rm Z})}$ are not.  The objects
\bq \hat{B}_{\mu\nu}^{({\rm Z})}=B_{\mu\nu}^{({\rm Z})}
    +\bpl V_{[\mu}V_{\nu]}^{(z)}\bpr^{({\rm Z}-1)}
 \eq
are, however, $\t$-invariant.  Equation (3) is the
${\rm Z}=1$ version of this
equation. The transformation rules for all ${\rm Z}\ge 1$ are given by
\brr \d\phi^{({\rm Z})} &=& \bar{\e}^i\l_i^{({\rm Z})}
     +\bar{\e}_i\l^{({\rm Z})i} \nonumber\\
     %%%
     \d V_\mu^{({\rm Z})} &=& i\ve^{ij}\bar{\e}_i\g_\mu
     \bpl 2X\l_j+\phi\O_j\bpr^{({\rm Z})}
     +i\bar{\e}^i D_\mu\l_i^{({\rm Z}-1)}+h.c. \nonumber\\
     %%%
     \d\hat{B}_{\mu\nu}^{({\rm Z})} &=&
     -\bar{\e}^i\s_{\mu\nu}
     \bpl 8\phi |X|^2 \l_i+4 \phi^2 \bar X \O_i
     \bpr^{({\rm Z})}
     +2i\ve^{ij}\bar{\e}_i\g_{[\mu}\bl
     V_{\nu]}^{(z)}\bpl 2X\l_j
     +\phi\O_j\bpr\br^{({\rm Z}-1)} \nonumber\\
     & & -\ve^{ij}\bar{\e}_i\g_{[\mu}D_{\nu]}
     \bpl 4\phi X\l_j+\phi^2 \O_j\bpr^{({\rm Z}-1)}
     +i \bar{\e}^i\bpl\l_i F_{\mu\nu}\bpr^{({\rm Z-1})}+h.c. \nonumber\\
     %%%
     \d\l_i^{({\rm Z})} &=& \bpl\Dslash\phi^{({\rm Z})}
     -i\Vslash^{({\rm Z}+1)}\bpr\e_i
     -\frac{i}{2X}\ve_{ij}\s\cdot\big(F-i\phi\F\big)^{(Z)}\e^j
     +\big(2\ve_{ij}\bar{X}\phi^{(Z+1)}
     -\frac{\phi^{(Z)}}{2X}Y_{ij}\big)\e^j\nonumber\\
     & & +\frac{1}{X}\bl
     \frac{1}{2\phi}\e^j\big(X\,\bar\l_i\l_j
     -\bar{X}\,\ve_{ik}\ve_{jl}\bar\l^k\l^l\big)
     -\big(\bar\e^j\O_j\big)\l_i-\big(\bar\e^j\l_j\big)\O_i\br^{(Z)}
 \label{rules}\err
Aside from completeness, we include these details to indicate
a new feature which is present in the general transformation
rules.  At all levels ${\rm Z}\ge 1$, a supersymmetry
transformation involves objects both at the next higher level,
${\rm Z}+1$, and also at the preceding level, ${\rm Z}-1$.
This is to be compared
with the case of the hypermultiplet, which also involves a
central charge hierarchy.  In that case
supersymmetry transformations link only to objects at higher levels
${\rm Z}+1$; they do not involve ${\rm Z}-1$.
Notice that the transformation rules for $\phi^{(Z)}$ and
$\l_i^{(Z)}$ are simply $z$-transformed versions of the
corresponding rules for $\phi$ and $\l_i$. This reflects
the fact that central charge transformations and supersymmetry
transformations commute when applied to these objects.
In contrast, the transformation rules for
$V_\mu^{(Z)}$ and $\hat{B}_{\mu\nu}^{(Z)}$
do not share this property with their $Z=0$ analogs.
This reflects the fact that $V_\mu$ and $B_{\mu\nu}$ are gauge
fields and that central charge transformations and
supersymmetry transformations commute into field-dependent
vector and tensor gauge transformations when acting on them.
One may wonder, since there
exists an infinite sequence of multiplets which are
joined above and below by supersymmetry, how it can be that a
{\it lowest} multiplet exists.  The answer to this seeming paradox
is that, as made clear above, for the lowest lying multiplet,
the vector and tensor fields are the gauge fields associated with
certain symmetries. The respective gauge transformations then appear in the
algebra in place of a linkage to a lower lying multiplet.
We stress this point especially; the lowest lying multiplet in the
central charge hierarchy is special in this respect.

Using the components of the vector-tensor multiplet
$(\phi,V_\mu,B_{\mu\nu},\l_i,\phi^{(z)})$ and the background
vector multiplet $(X,\O_i,W_\mu,Y_{ij})$, we construct a
linear multiplet $(L_{ij},\vphi^i, G, E_\mu)$ by requiring
the lowest component $L_{ij}$ to have weights $w=2$ and $c=0$ and to
transform into a spinor doublet
$\varphi_i$ according to $\d L_{ij}= \bar \e_{(i}\varphi_{j)} +
\varepsilon_{ik}\varepsilon_{jl} \bar \e^{(k}\varphi^{l)}$.
The expression for $L_{ij}$ is an extension of the linearized result
presented in \cite{DWKLL} and is given by
\bq L_{ij} = \phi\bpl X\,\bar{\l}_i\l_j
     +\bar{X}\,\ve_{ik}\ve_{jl}
     \bar{\l}^k\l^l\bpr
     +\frac{1}{3}\phi^3 \,Y_{ij}\,.
 \eq
We should point out that the
existence of such a multiplet in the vector multiplet background
depends sensitively on the specific nonlinear
transformation rules given above. The higher components of the
linear multiplet are constructed by successive supersymmetry
transformations. We refrain {}from giving these expressions here,
which are complicated but straightforward to compute.
{}From the product of a vector and a linear
multiplet one can construct an invariant action as described in
\cite{DWVHVP,DWLVP}.  The vector multiplet in this construction
must coincide with the vector multiplet that gauges the central
charge. In this way we arrive at the Lagrangian
\brr {\cal L} &=& 2|X|^2\phi\, \big(D^\mu\phi\big)^2
     -\ft{2}{3}\phi^3
     \bpl X\Box\bar{X}+\bar{X}\Box X\bpr \nonumber\\
     & & -2|X|^2\phi\, \big(V_\mu^{(z)}\big)^2
     +\ft{1}{4}\phi\,
     \bpl F_{\mu\nu}+i\phi\tilde{\F}_{\mu\nu}\bpr^2\nonumber\\
     & & -8|X|^4\phi\,(\phi^{(z)})^2
     +\ft{1}{12}\phi^3 \,Y_{ij}Y^{ij} \nonumber\\
     & & +4|X|^2\phi\,\phi^{(z)} W^\mu D_\mu\phi
     +\ft{1}{3}\phi^3 \,
     W_\mu\,\der_\nu\F^{\mu\nu}\nonumber\\
     & & +\phi\, W_\mu \bpl V^{(z)}_\nu
    (F^{\mu\nu}+i\phi\tilde{\F}^{\mu\nu}) -i D_\nu\phi\,
    (\tilde{F}^{\mu\nu}+i\phi\F^{\mu\nu})\bpr \nonumber\\
     & & +{\rm fermion\, terms} \,.
  \label{vtaction}\err
Equation (\ref{vtaction}) describes a supersymmetric action
involving the fields of the vector-tensor multiplet which is also
invariant under a local central charge transformation.
Note that the explicit factors of $W_\mu$ ensure that ${\cal L}$ transforms,
under the central charge, into a total derivative.

As mentioned above, it is possible to convert this Lagrangian to a
(classically equivalent) Lagrangian involving two vector
multiplets.  Such a duality transformation
is performed, in the usual manner,
by introducing a Lagrange multiplier field $a$, which, upon integration, would
enforce the Bianchi identity, shown in (\ref{Bianchi}),
imposed on the field strength $H^\mu$,
\bq {\cal L}_{\rm multiplier} = a\Big(D_\mu H^\mu
    +\ft{i}{4}F_{\mu\nu}\tilde F^{\mu\nu}
    +\ft{i}{2} \hat B_{\mu\nu}^{(z)}\tilde{\cal F}^{\mu\nu}\Big)\,.
 \eq
Including this term, we treat $H^\mu$ as unconstrained and
integrate it out of the action.  In equation (\ref{vtaction})
the $H^\mu$ dependence is implicit in $V_\mu^{(z)}$ via
the first equation of (\ref{central}).  In terms of $V_\mu^{(z)}$, the
equation of motion for $H^\mu$ takes on a simple form,
given by
\bq V_\mu^{(z)}=\der_\mu a.
 \label{vvac}\eq
The natural gauge fields in the dual theory are
found to be $W_\mu^0=W_\mu$ and $W_\mu^1=V_\mu+aW_\mu$, which
transform under a combined  central charge and gauge
transformation as $\d W_\mu^0=\der_\mu z$ and
$\d W_\mu^1=\der_\mu\big(\t+az\big)$.
Thus the enigmatic entagling of the gauge and central charge
transformations exhibited in (\ref{gauge})
satisfyingly disentangles in the dual formulation, leaving us with
an abelian gauge structure with field strengths
$\F_{\mu\nu}^I=2\der_{[\mu}W_{\nu]}^I$.  The dual action
also involves two complex scalars,  $X^0=X$ and $X^1=X(a+i\phi)$.
One can verify that $X^1$ and $W^1_\m$ transform as the
scalar and gauge field of a single vector multiplet. This is
confirmed by rewriting the bosonic Lagrangian in the dual
formulation (integrating out auxiliary fields) in terms of the
derivatives of a holomorphic function $F(X^0,X^1)$,
\bq {\cal L} = \frac{i}{2}\big(\der_\mu F_I\,\der^\mu\bar{X}^I
     -\der_\mu X^I\,\der^\mu\bar{F}_I\big)
     -\frac{i}{8}\big(\bar{F}_{IJ}\,\F^{+I}_{\mu\nu}\F^{+\mu\nu J}
     -F_{IJ}\,\F^{-I}_{\mu\nu}\F^{-\mu\nu J}\big)\,,
 \label{aform}\eq
where a subscript $I$ denotes differentiation with respect to
$X^I$. This is the generic form for the $N=2$ supersymmetric
action of vector multiplets. The function in the case at hand is
found to be
\bq F(X^0,X^1)  = -\frac{1}{3}\frac{(X^1)^3}{X^0}\,.
 \label{fff} \eq

Prior to performing the duality transformation, we could have
included $n-1$ additional vector multiplets, labeled by $I=2,
\cdots ,n$. The coupling of these vector multiplets would in
principle involve the background vector multiplet as well and would be
characterized by another holomorphic function involving
$X^0, X^2,...,X^n$. The dual Lagrangian would be
encoded in a function which is the {\it sum} of this
new function and (\ref{fff}). Hence, in this extended formulation,
the appearence of $X^1$ would be strongly restricted.
In all such cases, the Lagrangian is invariant, up to a total
divergence,  under
\bq
X^1\to X^1 + b \,X^0\,,\qquad W^1_\m \to W^1_\mu + b\, W^0_\m\,,
\eq
where $b$ is an arbitrary real parameter. This is the generalized
Peccei-Quinn symmetry associated with the axion field $a$.

In order to make contact with $N=2$ heterotic string
compactifications, we must couple the above Lagrangian to
supergravity. As we have stressed, this coupling should follow
straightforwardly within our adopted strategy. Therefore it comes
as no surprise that the
function (\ref{fff}) is homogeneous of second degree, which is
precisely the condition that must be satisfied in order to couple
vector multiplets to supergravity. In fact, with
the supergravity couplings included, the theory
based on (\ref{fff}) coincides with the dimensional reduction of
pure five-dimensional supergravity.
Although we find no indication that the coupling of
supergravity will lead to surprises, our result is somewhat
unexpected from the point of view of string theory. By starting
from a vector-tensor multiplet, which is one of the supermultiplets of
vertex operators in the compactified heterotic string
(cf. the discussion in \cite{DWKLL}), one would expect the
dilaton field $S=\phi-ia$ to exhibit stringy features. This
expectation does not seem fulfilled, however, as the dilaton
coupling in our construction is not universal.
A vector-tensor multiplet seems unable to couple to vector
multiplets other
than the one associated with its central charge. Although we can
arrange that the dilaton is subject to an $SU(1,1)$
$S$-duality invariance, the special K\"ahler space does not
factorize, so our solution does not meet the conditions
of the theorem of \cite{FVP}. More firm conclusions regarding
this issue necessitate further work on the supergravity coupling.
The results reported here are a first step in that direction.

\vspace{1cm}

We acknowledge useful discussions with J. Louis, D. L\"ust and A.
Van Proeyen. \\
For two of us (M.F. and B.K.) this investigation was performed as part of
the research program of the ``Stichting voor Fundamental Onderzoek
der Materie" (F.O.M.).\\
R.S. would like to thank the Belgian IISN for financial support.\\
This work was carried out in the framework of the European
Community Research Programme ``Gauge Theories, Applied
Supersymmetry and Quantum Gravity'', with a financial
contribution under contract SC1-CT92-0789.


\begin{thebibliography}{99}
\bibitem{grimm} R. Grimm, M. Sohnius and J. Wess,
Nucl. Phys. {\bf B133} (1978) 275.
\bibitem{sohnius} M.F. Sohnius, K.S. Stelle and P.C. West, Phys.
Lett. {\bf B92} (1980) 123.
\bibitem{DWKLL} B. de Wit, V. Kaplunovsky, J. Louis and D. L\"ust,
Nucl. Phys. {\bf B451} (1995) 53 (hep-th/9504006).
\bibitem{fayet} P. Fayet,
Nucl. Phys. {\bf B113} (1976) 135.
\bibitem{DWVH}  B. de Wit and J.W. van Holten,
Nucl. Phys. {\bf B155} (1979) 530.
\bibitem{siegel} W. Siegel and N. Berkovits, preprint
IFUSP-P-1180, ITP-SB-95-41, hep-th/9510106 \\
W. Siegel, preprint ITP-SB-42, hep-th/9510150.
\bibitem{zachos} C. Zachos,
Phys. Lett. {\bf 76B} (1978) 329.
\bibitem{DWVHVP}  B. de Wit, J.W. van Holten and A. Van Proeyen,
Phys. Lett. {\bf 95B} (1980) 51.
\bibitem{DWLVP} B. de Wit, J.W. van Holten and A. Van Proeyen,
Nucl. Phys. {\bf B184} (1981) 77 \\
B. de Wit, P.G. Lauwers and A. Van Proeyen,
Nucl. Phys. {\bf B255} (1985) 569.
\bibitem{FVP} S. Ferrara and A. Van Proeyen,
Class. Quantum Grav. {\bf 6} (1989) L243.
\end{thebibliography}
\end{document}